\documentclass[journal]{IEEEtran}
\usepackage{graphicx}
\usepackage{cite}
\usepackage{hyperref}
\usepackage{amsmath}
\usepackage{array,graphicx}
\usepackage{multirow}
\usepackage{kantlipsum}
\usepackage{amssymb}
\usepackage{float}
\usepackage{booktabs}
\usepackage{adjustbox}

\newif\ifIEEEpublication
\IEEEpublicationfalse 

\begin{document}

\title{Adaptive Coordination Offsets for Signalized Arterial Intersections using Deep Reinforcement Learning}

\author{\IEEEauthorblockN{Keith Anshilo~Diaz\P, Damian~Dailisan\P, Umang~Sharaf, Carissa~Santos, Qijian~Gan, Francis~Aldrine~Uy, May~T.~Lim, and~Alexandre~M.~Bayen,~\IEEEmembership{Member,~IEEE}}
\thanks{Manuscript received ... This work was funded by the Commission on Higher Education: Philippine--California Advanced Research Institutes (CHED--PCARI grant IIID-2016-006).}
\thanks{\P These authors contributed equally to this work.}
\thanks{K.A.L. Diaz, C. Santos, and F.A. Uy are with the School of Civil, Environmental and Geological Engineering, Mapua University, Manila, Philippines (e-mail: \{kaldiaz,faauy\}@mapua.edu.ph,cjrnsantos@gmail.com)}%
\thanks{D. Dailisan and M. T. Lim were with the National Institute of Physics, University of the Philippines Diliman (e-mail: \{ddailisan,may\}@nip.upd.edu.ph). D. Dailisan is now currently with the Analytics, Computing, and Complex Systems Laboratory, Asian Institute of Management, Philuppines.}%
\thanks{U. Sharaf was with the Electrical Engineering and Computer Sciences department at University of California, Berkeley (e-mail: umang@berkeley.edu)}%
\thanks{Q. Gan is with the PATH program in the Institute of Transportation Studies, University of California, Berkeley (e-mail: qgan@berkeley.edu)}%
\thanks{A. M. Bayen is with the Electrical Engineering and Computer Sciences Department, also with the Department of Civil and Environmental Engineering, and also with the Institute of Transportation Studies, University of California, Berkeley (e-mail: bayen@berkeley.edu)}
}


\maketitle

\begin{abstract}
Coordinating intersections in arterial networks is critical to the performance of urban transportation systems.
Deep reinforcement learning (RL) has gained traction in traffic control research along with data-driven approaches for traffic control systems.
To date, proposed deep RL-based traffic schemes control phase activation or duration. 
Yet, such approaches may bypass low-volume links for several cycles to optimize the network-level traffic flow.
Here we propose a deep RL framework that dynamically adjusts offsets based on traffic states and preserves the planned phase timings and order derived from model-based methods.
This framework improves arterial coordination while maintaining phase order and timing predictability.
Using a validated and calibrated traffic model, we trained the policy of a deep RL agent that aims to reduce travel delays in the network.
To evaluate the resulting policy, we compared its performance against the phase offsets deployed along a segment of Huntington Drive in the city of Arcadia.
The resulting policy dynamically readjusts phase offsets in response to changes in traffic demand.
Simulation results show that the proposed deep RL agent outperformed the baseline on average, effectively reducing delay time by 13.21\% in the AM Scenario, 2.42\% in the Noon scenario, and 6.2\% in the PM scenario when offsets are adjusted in 15-minute intervals.
Finally, we also show the robustness of our agent to extreme traffic conditions, such as demand surges in off-peak hours and localized traffic incidents.
\end{abstract}


\IEEEpeerreviewmaketitle

\section{Introduction}
\IEEEPARstart{T}{raffic} congestion remains one of the most complex problems faced by cities worldwide.
An average driver in Los Angeles spent up to 11\% of their day in traffic congestion, which cost the US economy \$87 billion in 2018~\cite{cookson2017inrix}.
As the demand for transportation increases in urban areas, there is a need to develop systems that help manage traffic congestion.
Arterial roads, in particular, suffer from congestion, and much of the work to optimize arterial networks focuses on the design of intelligent systems that adapt to highly dynamic traffic demands~\cite{ezell2010explaining}.
The most common approach to improve congestion in arterial roads uses systems that \emph{coordinate} signals between neighbor intersections to improve traffic flow.
Some of the well-known control strategies are MAXBAND~\cite{little1966synchronization}, MULTIBAND~\cite{gartner1991multi}, and TRANSYT~\cite{robertson1969transyt}. 
Model-based traffic control systems typically optimize for a particular demand scenario;
multiple control plans with different phase offsets
account for the variation of demand throughout the day~\cite{wang2018review, roess2004traffic,zou2004timing}.
These algorithms only optimize for particular directions, a limitation that can result in congestion in collector roads and other parts of the network~\cite{gettman2007data}.
Systems that optimize a static objective function using averaged/aggregated traffic measures (rather than learned from live data) have limited robustness in handling highly dynamic conditions as well as exogenous uncertainties (e.g. traffic incidents, demand surges) ~\cite{rodrigues2019towards}. 

Advances in computing power and the availability of large data sources have driven new perspectives in solving problems.
In 2015, breakthroughs in deep reinforcement learning (RL) led to superhuman mastery in playing Atari Games~\cite{mnih2015human} and Go~\cite{silver2017mastering}.
Deep RL has the potential to dynamically optimize traffic controllers in ways that are difficult to attain using traditional control.
Current works have agents that control phases activation or duration~\cite{Wei2018intellilight, Wei2019presslight, Wei2019colight} or use deep RL to solve control systems modeled by non-linear partial differential equations~\cite{belletti2017expert}.
But these approaches can lead to situations wherein low volume links are bypassed for multiple cycles, which is undesirable for affected drivers.

Our work leverages the potential of deep RL to coordinate a series of signalized intersections along an arterial corridor to dynamically adjust phase offsets to reduce travel delays in the network while preserving the phase order and timings used in the field.
We chose to focus solely on control of the phase offsets, which avoids long wait times for drivers on minor roads.
Our work also used an arterial traffic model developed in the Connected Corridors I-210 Pilot that is calibrated and validated by field data~\cite{dion2015connected,gan2017estimation,gan2019arterial,bayen2011mobile}.
This results in a policy that dynamically adjusts phase offsets and is robust enough to adapt to varying traffic demand scenarios.
We retained model-based phase splits while using a model-free method for coordination, which, to the best of our knowledge, is a first among deep RL control variations.
Our proposed approach does not seek to overhaul the current state of practice but instead rests in the middle ground between existing model-based traffic control methods and model-free deep RL.


This article discusses our work on using deep RL to achieve adaptive coordination of signalized arterial intersections.
Section \ref{prelim} presents the background of deep RL and the essential parameters of RL formalization.
Section \ref{method} discusses our proposed parameterization of the RL problem, including the choice of detector measurements for observations and designing the reward function.
We also detail our chosen training configuration, learning algorithm, and traffic network model.
Section \ref{results} provides the results and discussions.
Lastly, we summarize our work in Section \ref{conclusion}.

\section{Related Work}
\label{rrl}
The advent of centralized computer-controlled traffic systems in the 1960s triggered a shift in the design of control algorithms from static~\cite{webster1958traffic} to  dynamic~\cite{dunne1964algorithm}. 
Traffic control optimizes competing streams of vehicles at an intersection to prevent bottlenecks due to high traffic demand.
This section discusses two ways to address the problem: model-based and the deep RL approach.

\subsection{Model-based coordination control}
Conventional traffic engineering methods for optimization minimize a cost function while assuming static averages as inputs~\cite{wei2019survey}.
In practice, the coordinated system for traffic signal control is most used to provide continuous progression to reduce delay time on arterial roads~\cite{roess2004traffic}. 
Short-term variations in traffic, however, make it challenging to use pre-planned offsets in field deployments~\cite{park2010quantifying}.
Various approaches for adaptive control algorithms try to account for the stochasticity of traffic~\cite{gettman2007data,coogan2017offset,yin2007offline}.
However, their generated offset values only provide best results “on average".
This further emphasizes the need for real-time offset adjustment that is adaptive to changes in traffic conditions~\cite{abbas2001real}.

\subsection{Deep Reinforcement Learning approach}
Deep RL has gained significant traction in the field of traffic control because of its goal-oriented approach in learning through trial-and-error in an environment~\cite{wu2017flow,vinitsky2018benchmarks}. 
It features intelligent agents that learn by repeated interactions with an environment~\cite{van2016coordinated} and provides a well-suited framework for optimization problems in transportation~\cite{schultz2018deep}.
Training the agent to interact with high-fidelity simulator environments such as Aimsun~\cite{Casas2010} and SUMO~\cite{SUMO2018} allows it to make decisions that optimize an objective and learn tasks such as traffic control coordination.

Early works that used RL for dynamic control of traffic lights used tabular Q-learning, which was limited to small state representations and discrete action spaces~\cite{el2014design}.
Recent developments in performance computing led to the use of deep RL for the traffic signal control problem~\cite{Wei2018intellilight,ge2019cooperative,chen2019adaptive} which greatly increased scalability and extent.
Groups of intersections can also learn cooperative control using \emph{Multi-agent Reinforcement Learning} (MARL).
Coordination requires shared actions and states between controllers to promote coordination~\cite{wiering2000multi,mikami1994genetic}, which Q-learning achieves through the transfer of Q-values of adjacent intersections to a network-level loss function~\cite{ge2019cooperative}.
MARL can leverage parallel computing for distributed learning, which gives the advantage of significantly reducing computing time compared to a centralized/single-agent RL approach.
While this approach reduces computing time, it does not guarantee global optimality if agents' actions are uncoordinated~\cite{bucsoniu2010multi}. 
Other works address this by allowing agents to partially observe other intersections, which provides agents a sense of states and actions of other intersections~\cite{steingrover2005reinforcement,el2013multiagent,houli2010multiobjective,dulac2015deep}, or by integrating a transport-based theory into the MARL algorithm~\cite{kuyer2008multiagent}.
These studies struggle to achieve fully coordinated agents, as they suffer from the curse of dimensionality.

Our work provides a centralized deep RL solution that is a more scalable, realistic, and practical approach for the traffic light coordination problem.
We limited agent actions to control of \emph{phase offsets} which avoids \emph{curse of dimensionality} for centralized deep-RL agents and preserves the signal timing plans of individual intersections.
This approach enables a single agent to control a group of subsequent intersections and, most importantly, leverage both model-based and model-free approaches.
Moreover, we used a well-calibrated microscopic model in the arterial network traffic simulation~\cite{gan2019arterial}.

\section{Reinforcement Learning Background}
\label{prelim}

Reinforcement Learning (RL) is a method of learning through interactions between an agent and the environment.
Agent's actions are not predetermined but instead discovered by maximizing a reward function to reach the desired state~\cite{sutton1998introduction}.
A \emph{Markov Decision Process} (MDP)~\cite{bellman1957markovian} provides a mathematical discrete-time formulation of how RL learns decision-making processes in uncertain environments.
An MDP is a tuple $\langle \mathcal{S}, \mathcal{A}, R, P \rangle$ where $\mathcal{S} \subseteq \mathbb{R}^{n}$ is the set of all possible states of the environment (partially or fully observable) in $n$ dimensions, $\mathcal{A} \subseteq \mathbb{R}^{m}$ is the $m$-dimensional action space, $R \in (\mathbb{R}^n, \mathbb{R}^m) \to \mathbb{R}$ is the reward function that dictates the numerical reward for the state~$s^{\prime}$  observed by the agent after taking action~$a$ at state~$s$, and the transition probability function~${P}$.
Markov processes describe sequences of events where the probability move to the next state only depends on the present state and not on past states.

The goal of an agent is to find an optimal policy~$\pi^*$ that maximizes the sum of discounted rewards
\begin{equation}R(\tau)=\sum_{t=0}^{H_{T}} \gamma^{t} r_{t}.\end{equation}
The horizon~$H_{T}$ is the number of intervals (which can include several simulation timesteps) that the agent performs actions.
The trajectory~$\tau$ represents a sequence of state-action pairs up to the horizon, and the discount factor $\gamma \in[0,1]$ adjusts the contribution of future states to the reward.
To find the optimal policy, we select the policy that maximizes the value or expected return, which is referred to as the value  and action-value functions~\cite{bellman1957markovian}, given by 
\begin{eqnarray}V^{\pi}(s)&=&\underset{a \sim \pi}{\mathbb{E}}\left[r(s, a)+\gamma V^{\pi}\left(s^{\prime}\right)\right],\label{eqn:bellman1}\\
Q^{\pi}(s, a)&=&\underset{s^{\prime} \sim P}{\mathbb{E}}\left[r(s, a)+\gamma \underset{a^{\prime} \sim \pi}{\mathbb{E}}\left[Q^{\pi}\left(s^{\prime}, a^{\prime}\right)\right]\right],
\label{eqn:bellman2}
\end{eqnarray}
where $x\sim y$ denotes $x$ is drawn from a distribution $y(x)$.
These equations estimate how good the current state is by considering the expected reward of all possible future states.

Deep RL refers to RL algorithms that approximate (\ref{eqn:bellman1}) and (\ref{eqn:bellman2}) using deep neural networks.
Unlike supervised learning, deep RL does not rely on labeled datasets; training data is generated from interactions of the agent's current policy with the environment.
RL algorithms used to estimate reward and action probabilities are classified into value-based, policy gradient, and actor-critic methods.
Our work uses \emph{Proximal Policy Optimization} (PPO)~\cite{schulman2017proximal}, a policy gradient method 
that encourages actions that are better than other actions on average. 
This method uses relative advantage to evaluate actions $a$ taken in state $s$ compared to other actions obtained from the current policy $\pi$.

\section{Methodology}
\label{method}
\begin{figure}[tb]
    \centering
    \includegraphics[width=70mm]{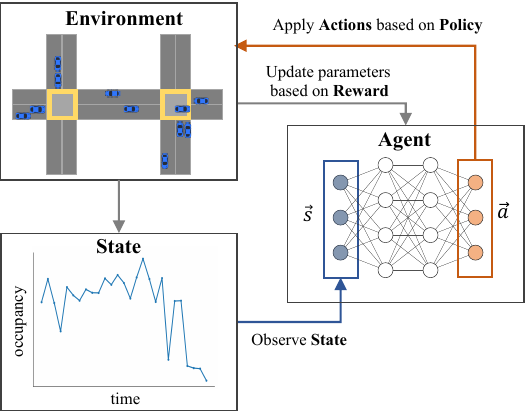}
    \caption{Deep RL Framework for Traffic Control Coordination}
    \label{fig:RLDiagram}
\end{figure}

This section describes the models used in our simulation and how we implemented deep RL to coordinate intersections through dynamic adjustments of phase offsets.
Figure \ref{fig:RLDiagram} depicts a schematic of using deep RL for traffic control.

\subsection{Traffic Control Coordination using Deep RL}

Traffic simulations are run using \emph{Flow}~\cite{wu2017flow} over an Aimsun-based traffic model.
For a system with $N_\mathrm{target}$ coordinated intersections, we create an RL agent with the following parameters:

\subsubsection{State Representation}
\begin{figure}[tb]
    \centering
    \includegraphics[width=2.7in]{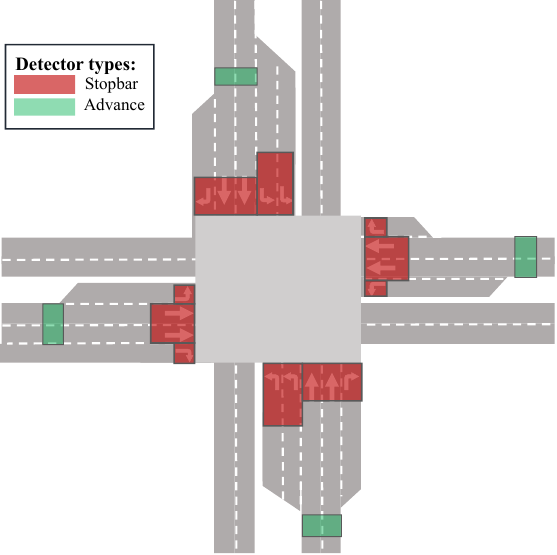}
    \caption{Detector layout for the intersection of Huntington and Santa Anita with advance (green) and stopbar (red) detectors (not to scale.)
    Detectors can cover multiple lanes, and reflect what is deployed in the field.
    Some links may have lanes that are uncovered by detectors, or in some cases, one or even both types of detectors.}
    \label{fig:det_lay}
\end{figure}

The locations of detectors in the simulation correspond to existing deployed detectors.
For each intersection, we assume 4 incoming links, each with 2 types of detectors: \emph{advance} and \emph{stop bar} detectors (Fig. \ref{fig:det_lay}).

\begin{figure*}[t!b]
    \centering
    \includegraphics[width=140mm]{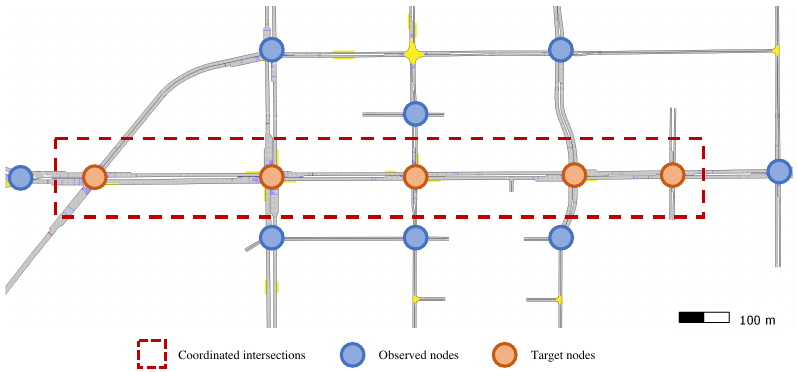}
    \caption{Huntington Drive in Arcadia, CA. Target nodes are the five coordinated intersections, and observed nodes are first degree neighbors of target nodes. Leftmost node is the first intersection, and the rightmost node is the fifth intersection.}
    \label{fig:Network}
\end{figure*}
Flow and occupancy measures detected by each detector type capture different aspects of traffic dynamics~\cite{gan2019arterial}.
Flow--occupancy measurements from advance detectors reflect accurate flow and occupancy measures on the link.
On the other hand, measurements from stop bar detectors are sensitive to signal control and downstream conditions such as queue spillback.
We combine the information captured by both of these detectors to enrich the state space for training the agent.

We encoded the partially observed state of the network as an $(N_\mathrm{target} + N_\mathrm{observed})\times 4\times 2\times 2$ dimensional vector where $N_\mathrm{target}$ represents the target nodes corresponding to the coordinated intersections and $N_\mathrm{observed}$ represents nodes that are first-degree neighbors of the target nodes (Fig. \ref{fig:Network}).
In cases where links have multiple detectors (Fig. \ref{fig:det_lay}), the measurements from the same detector type (\emph{advance} or \emph{stopbar}) are aggregated for each link.
When a detector is not present for a link, its corresponding element in the state vector is set to zero.

In a 4-hour simulation period, we collected sensor measurements over an interval of $\tau=15\,\mathrm{minutes}$.
An episode of training consists of 3 runs of a 4-hour simulation.
Instead of training against single time periods, we randomly select among the three time periods (AM, Noon, and PM) for each runs.
This choice exposes the agent to varying traffic demands while avoiding overfitting to a particular sequence of traffic conditions (AM, Noon, PM), and greatly increases the robustness of the learned policy.

\subsubsection{Action Space}\label{sec:actions}
Our proposed approach preserves the existing order of signal phases.
The agent attempts to learn optimal offsets between intersection timings.
Actions are $N_\mathrm{target}$ integer offsets from  $t_\mathrm{offset}\in\left[0,t_c\right)$, where $t_c$ is the cycle period, which is the same for all intersections. The agent adjusts the offset every 15 minutes.
To adjust offsets the current phase is extended (or reduced) by  $t_\mathrm{offset}^\mathrm{new}-t_\mathrm{offset}^\mathrm{current}$, or $(t_\mathrm{offset}^\mathrm{new}-t_\mathrm{offset}^\mathrm{current})-t_c$ when the difference exceeds $t_c/2$.
If the current phase is shorter than the needed reduction, it is skipped and the next phase is shortened accordingly.

\subsubsection{Reward Function}
Intuitively, a good traffic policy minimizes travel time.
However, when designing RL experiments, measuring individual travel times is costly, as the simulator would need to keep track of each vehicle's travel times. 
Zheng et al. showed the proportionality between queue lengths and travel time for a single intersection, with $r_t=-\sum_{i}l_{i,t}$~\cite{Zheng2019}:
\begin{equation}
    \bar{T} = \frac{\tau \times \bar{l}}{N} + \frac{L}{\mu},
    \label{eq:queue_travel_time}
\end{equation}
where $\bar{l}$ is the average queue length of all lanes in the intersection, $\bar{T}$ is the average travel time of all vehicles, $\tau$ is the measurement time interval, $N$ is the number of vehicles present, $L$ is the length of the section, and $\mu$ is the free-flow speed of a vehicle.
This expression shows that minimizing queue lengths for a single intersection is equivalent to minimizing travel times.
In another work, Gan et al. showed a linear relationship between travel times and vehicle queues for the same intersection corridor as our study site \cite{gan2019arterial}.
Moreover, using queue lengths for the reward function will give the agent a sense of the standing queues that hampers the progression of the platoon.

We modify Zheng's reward function to take into account intersections with varying road lengths by rescaling with the length of the section; our reward is $r_t =-\sum_{i} l_{i,t} \times\left(\frac{c}{L_{i}}\right)$ where the queue length~$l_i$ is scaled by the average car length~${c}$ (which is set to 5 meters) and the length of the section ~${L_{i}}$, for all $i$ incoming links at all intersections from $N_{t}$ and $N_{o}$.
When all section lengths~$L_i$ are the same for all intersections, an additional factor of $c/L_i$ appears at the first term of (\ref{eq:queue_travel_time}).
First-degree neighbors of the coordinated intersections are considered in the reward function for the agent to get a sense of balancing the queues in the whole network.

Expressing the reward as the sum of scaled queue lengths has two implications.
First, scaling the rewards eliminates bias for major roadways/movements -- which generally have longer links -- that can have long queues.
Second, scaling the queue lengths also serves to normalize and control the gradients, and allows us to relate the reward to the overall state of queues in the network.

\subsection{Training Configuration}
Most deep RL applications for traffic control use \emph{deep Q-Networks}~(DQN). 
In general, Q-learning based DQN models are not suitable for complicated systems that have vast state and action spaces~\cite{haydari2020deep}.
For our work, we used PPO for its effectiveness in high-dimensional action and state spaces, which is not in the case for DQN.
More importantly, PPO performs at-par or better than state-of-the-art approaches with easier tuning and implementation~\cite{schulman2017proximal}.

We modify action probabilities by directly updating our policy network.
We used $64 \times 64 \times 64$ fully-connected hidden layers for the policy neural network.
To estimate the advantage function relative to the value function, we use a \emph{Generalized Advantage Estimator}~(GAE)~\cite{Schulman}.
The advantage estimate is the difference between the actual discounted reward and the expected reward based on the value function.
In PPO, the advantage~$A^{\pi}(s,a)$ is calculated after the episode sequence is collected from the environment. Hence, the discounted reward at each time step is known. The value function serves as the baseline, which estimates the final return in each episode.

Each of the three time periods, AM, Noon, and PM, is trained for 4 hours, and the remaining time is used to test the agent's flexibility. Training starts after a 15-minute warm-up period to ensure that there are vehicles on the road network at the start of the simulation.
Moreover, to ensure that the trained agent does not overfit to the three scenarios, the set of initial simulation seeds is different from the seeds used in the agent's evaluation. 
The full set of training hyperparameters used in this work are presented in Table \ref{table:hyperparams}.

\subsubsection{Learning Parameters}
The PPO algorithm trains a stochastic neural network policy~$\pi$ by directly updating probabilities.
To be specific, PPO uses two policies: the first is the current policy $\pi_{\theta}(a | s)$  that we want to update, and the second is the last policy that we use to collect samples $\pi_{\theta_k}(a | s)$.
The weights of the neural network are updated according to
\begin{equation}
\theta_{k+1}=\arg \max _{\theta} \underset{s, a \sim \pi_{\theta_{k}}}{\mathbb{E}}\left[L\left(s, a, \theta_{k}, \theta\right)\right],
\end{equation}
where $\theta_{k}$ and $\theta$ represent the weights from the old policy and current policy, respectively.

We used a variant of PPO that lacks a KL-divergence term in the objective function and instead relies on a clipping hyperparameter $\epsilon$, to limit the change between the new and old policies.
The objective function is
\begin{multline}
L\left(s, a, \theta_{k}, \theta\right)= \\
\min \left(\frac{\pi_{\theta}(a | s)}{\pi_{\theta_{k}}(a | s)} A^{\pi_{\theta_{k}}}(s, a),\quad g\left(\epsilon, A^{\pi_{\theta_{k}}}(s, a)\right)\right)\label{eqn:objective}
\end{multline}
The first term in (\ref{eqn:objective}) is the objective for normal policy-gradient methods that pushes the policy towards actions with greater advantage than the baseline. 
The second term $g(\epsilon, A)$ depends on the sign of the advantage: if the advantage is positive, then $g(\epsilon, A)=(1+\epsilon) A$; otherwise it is $g(\epsilon, A)=(1-\epsilon) A$.
Directly including the clipping operation in the objective function enables the algorithm to remove the incentive for drastic policy updates, thereby regularizing the policy by preventing updates that result in large deviations from the old policy.
The hyperparameter $\epsilon$ limits the magnitude of policy updates and prevents large deviations from the old policy.

\begin{table}[tb]
\centering
\caption{Training Hyperparameters}
\begin{tabular}{l l} 
Hyperparameter & Value \\
\toprule
Neural Network shape & $64 \times 64 \times 64$ (3 hidden layers) \\
Horizon, $H_{T}$ & 45 steps (12 hours in simulation) \\
Learning Rate & $5 \times 10^{-4}$ \\
Training batch size & 300 \\
SGD Minibatch size & 15 steps (4 hours in simulation) \\
Discount $(\gamma)$ & 0.99 \\
GAE parameter $(\lambda)$ & 0.97 \\
Clipping hyperparameter, $\epsilon$ & 0.2 \\
\end{tabular}
\label{table:hyperparams}
\end{table}

\subsection{Traffic Model}
\subsubsection{Arterial Network}
Realistic and robust simulation environments are vital components of RL;  learning relies primarily on interactions within the environment.
We chose a portion of Huntington Drive in Arcadia, California, a  busy arterial segment near I-210, as our test site.
This test site uses a portion of the I-210 Pilot that uses an Aimsun model developed by \emph{California Partners for Advanced Transportation Technology}~(PATH) at UC Berkeley under the Connected Corridors Program~\cite{dion2015connected}.
The model incorporates the actual timing plans used by the Transportation Management Center (TMC) in the city of  Arcadia.
The Connected Corridors project team has carefully calibrated the model with field data to mimic real-world traffic conditions. 
The network contains 34 nodes and 118 arterial road sections and has a total length of 14 km.

For the experiment, we chose $N_\mathrm{target}=5$ consecutive intersections with collector roads along Huntington Drive: Santa Clara Street, Santa Anita Avenue, First Avenue, Second Avenue, and Gateway Drive.
These five intersections on Huntington Drive are the target coordinated intersections; the agent obtained additional detector measurements from the target nodes' first-degree neighbors ($N_\mathrm{observed}=8$ nodes, see Fig. \ref{fig:Network}).
Cycle lengths for these intersections vary throughout the day (from 90 to 120 seconds, see Table \ref{table:sychoffsets}) to match the demand patterns.
At any given time, however, the timing plans for these intersections all have equal cycle lengths.
While offsets are normally constrained to be less than the cycle length, the action space must be fixed throughout training.
We chose our action space to have a maximum value of $t_c=120\textrm{\,sec}$, which is the longest cycle length used by the timing plans to accommodate the various cycle lengths used in the model.

\subsubsection{Traffic Demand}
\begin{figure}[tb]
    \centering
    \includegraphics[width=2.8in]{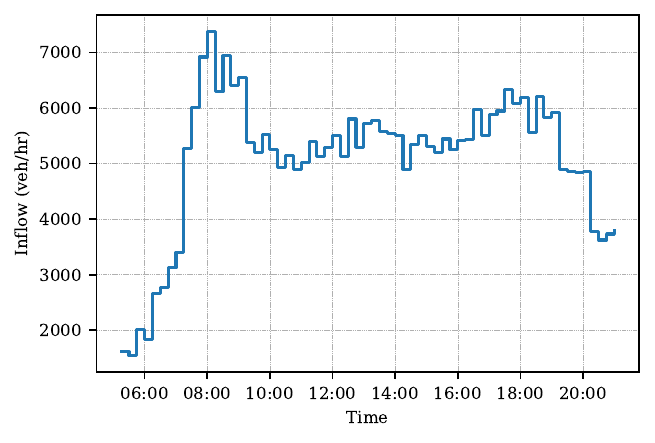}
    \caption{Flow of vehicles that pass through the coordinated intersections measured in 15 min. intervals across three time periods: AM (5:00 -- 10:00\,AM), Noon (10:00\,AM -- 2:00\,PM),  and PM (2:00 -- 9:00\,PM). We only use the first 4 hours of the AM and PM scenarios to train the deep RL agent.}
    \label{fig:inflows}
\end{figure}

The traffic estimates in the arterial model that we used are validated and calibrated with field detectors and signal phasing data~\cite{gan2019arterial}.
We used three demand scenarios (AM, Noon, and PM) so that the agent gets a sense of the different traffic demand dynamics throughout the day.
We measured the inflows for all links that will traverse our coordinated intersections (Fig. \ref{fig:inflows}) and observed three distinct features for the three scenarios: a surge in demand for the AM scenario, near-uniform demand during the Noon scenario, and reduced demand during the PM scenario.
We trained the agent only on the first four hours of the three different scenarios.

Exposing the deep RL agent to different demand profiles is a key feature of this work.
This sets our deep RL approach apart from other studies that use synthetic traffic demand; the demand that we used has been validated and calibrated against field sensor measurements.

\subsection{Performance Baseline}
To benchmark the RL agent, we compared its performance against traffic signals used in Huntington Drive in the city of Arcadia, CA, which is a part of the I-210 Connected Corridors Pilot. The Connected Corridors team coded these timing plans into an Aimsun model, using the data provided by the city of Arcadia. These timing plans have been optimized by SYNCHRO studies and field experiments before deployment to the field.

While our reward function used scaled queue lengths in training the RL agent, we used delays as a second metric to evaluate performance and further validate our results.
Delay is a widely used traffic measurement used to determine the Level of Service.
While we could have designed the reward to use delay, delay measurements need to track all vehicle trips, which require global (or at least, network-level) knowledge of vehicle travel times.
Queues, on the other hand, are local (intersection-level) measures.
Thus queues provide a convenient, measurable metric that can be adopted for training using simulation and used on targeted intersections in the field.

Using Aimsun, we measured the average delay for all vehicles in the system as 
\begin{equation}
    \mathrm{delay}=\frac{\sum_{i} T_i-T_{i,\mathit{free}}}{\sum_i \mathrm{(route\ distance)}_i},
\end{equation}
where the summation is carried out over all vehicles, $T_i$ is the travel time of the $i^{th}$ vehicle, and $T_{i,\mathit{free}}$ is the free flow travel time.
Scaling the delay in this way allows us to compare vehicle trips on routes with different distances.
We evaluated both our deep RL approach and the timing plans implemented in the field as our baseline by comparing the average delay (sec/km) of vehicles traversing the network.
We also tested the agent against exogenous uncertainties~\cite{rodrigues2019towards} such as demand surges and traffic incidents. Table \ref{table:sychoffsets} shows the schedule of baseline offsets.

\begin{table}[tb]
\caption{Baseline offsets for each intersection and cycle lengths (in seconds) for (a) AM and (b) Noon (c) PM scenarios}
\centering
\resizebox{3.4in}{!}{%
\begin{tabular}[width=3.4in]{c c c c c c c} 
Time              & Intersection & Intersection & Intersection & Intersection & Intersection & Cycle \\
              &  1 &  2 &  3 &  4 &  5 & Length\\
\toprule
5:00\,AM -- 5:45\,AM & 75       & 66       & 14       & 19       & 48  &  110   \\
5:45\,AM -- 6:30\,AM & 40       & 40       & 5        & 0        & 5    & 90    \\
6:30\,AM -- 9:00\,AM & 60       & 60       & 65       & 75       & 5    & 120    \\ 
9:00\,AM -- 11:00\,AM & 40       & 40       & 5        & 0        & 5   &  90   \\  \midrule
10:00\,AM -- 11:30\,AM & 40       & 40       & 5        & 0        & 5  &   90   \\
11:30\,AM -- 2:00\,PM  & 0        & 0        & 55       & 55       & 55 &  105    \\ \midrule
2:00\,PM -- 4:00\,PM & 0        & 0        & 55       & 55       & 55   &  105  \\
4:00\,PM -- 7:00\,PM & 60       & 60       & 65       & 75       & 5    &  120  \\
7:00\,PM -- 8:30\,PM & 0        & 0        & 55       & 55       & 55   &  105  \\
8:30\,PM -- 9:00\,PM & 40       & 40       & 5        & 0        & 5    &  90  \\
\end{tabular}
}

\label{table:sychoffsets} 
\end{table}

\section{Experimental Results and Discussion}
\label{results}

\subsection{Learning Progress}

\begin{figure}[tb]
    \centering
    \includegraphics[width=3.3in]{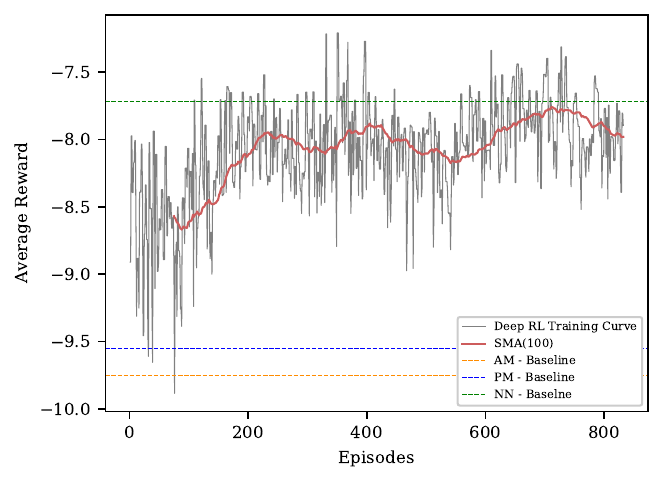}
    \caption{Rewards over episodes.
    Dashed lines represent the expected rewards of the three baseline scenarios.}
    \label{fig:rewards}
\end{figure}

We measured the expected rewards from the baseline scenarios and compared this against the training curves for deep RL (Fig. \ref{fig:rewards}).
Although our actions are sparse, the 100-episode moving average of the reward suggests that the agent is learning.
During training, the agent randomly picks one of the three scenarios (AM, Noon, and PM) with uniform probability; this tells us that the average baseline reward for the three scenarios is roughly -9.01, but the RL rewards of the agent exceed this value.
As a result, allowing the agent to experience the three scenarios (with different demand patterns and paths) during training lets us have a single trained agent with the combined traffic patterns in the three scenarios.
Deep RL algorithms, especially PPO, are computationally demanding in nature. Despite ending the training only after less than a thousand episodes, we can show that our deep RL agent outperforms timing plans in the field.

\subsection{Evaluation of Network Performance}

\begin{figure}[tb]
    \centering
    \includegraphics[width=3.2in]{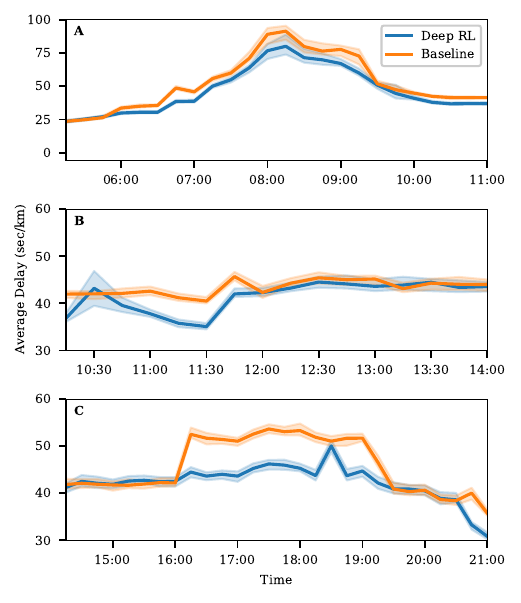}
    \caption{
    Comparison of performance of the offset adjustment of policy deep RL vs. baseline offsets show a
    reduced delay for scenarios a) AM scenario, b) Noon scenario and c) PM Scenario.}
    \label{fig:AM_average}
\end{figure}

We evaluate the performance of the deep RL agent against the baseline case by comparing the average delay of vehicles traversing the network (Fig. \ref{fig:AM_average}).
Seventy realizations of the model were used for each time period (AM, Noon, and PM), and the bands represent the $1^\mathrm{st}$ and $3^\mathrm{rd}$ quartiles of the average delay values.
Using deep RL to control offsets decreases the average delay by 13.21\% in the AM scenario, 2.42\% in the Noon scenario, and 6.2\% in the PM scenario compared to the baseline.

\begin{figure*}[tb]
    \centering
\includegraphics[width=6.5in]{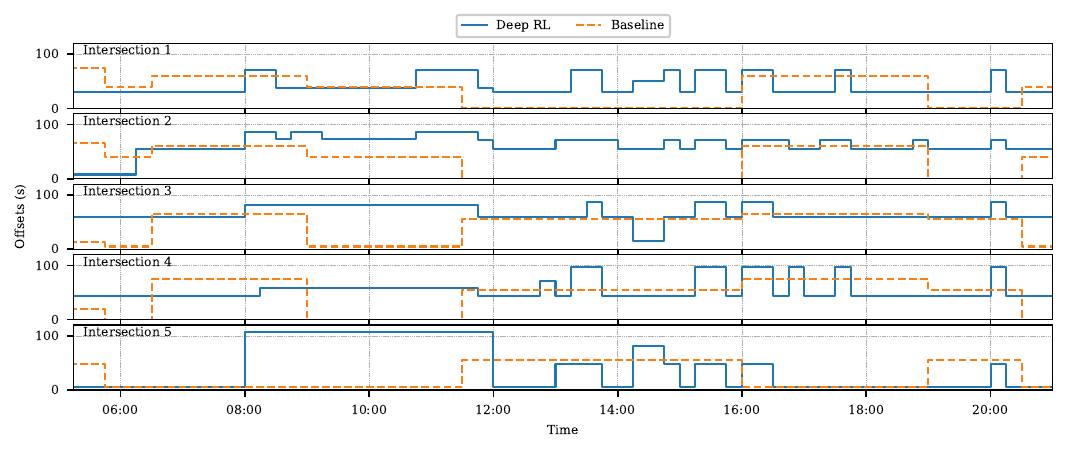}
    \caption{Time evolution of deep RL and baseline offsets. Deep RL produce offsets that change more frequently than the baseline offsets.}
    \label{fig:offsetcomp}
\end{figure*}

We note that the deep RL agent is only trained on the first four hours of each scenario; the remaining time is used to test the agent's performance on unfamiliar time periods.
However, this does not apply to the Noon scenario, which has a duration of exactly four hours.
We find that the learned policy mostly outperforms the baseline in the AM scenario, even in time periods (10--11\,AM) outside of training (Fig. \ref{fig:AM_average}a).
Comparison of the phase offset adjustments also shows that deep RL changes offsets more frequently than the baseline timing plan does during peak periods (Fig. \ref{fig:offsetcomp}), which consequently induces less delay.
The baseline timings used offsets optimized for time periods that were empirically selected by traffic engineers.
We see that deep RL allows for shorter action intervals that result in a more dynamic coordinated control of these intersections.
Although our current deployment may generate some offsets longer than the actual cycle length, only the fifth intersection is affected (9\,AM to 12\,NN).
Our implementation of offset adjustment in Sec. \ref{sec:actions} avoids extending the current phase by an amount exceeding $t_c/2$ and circumvents this potential conflict.

\begin{figure}[tb]
    \centering
    \includegraphics[width=3in]{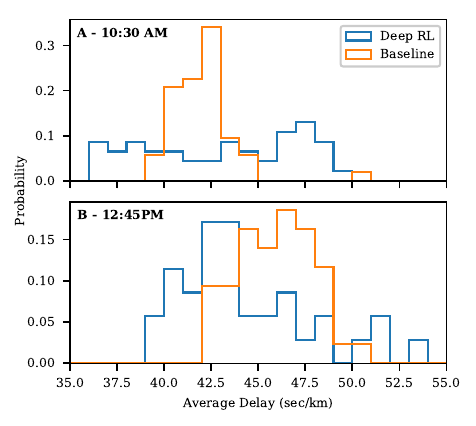}
    \caption{Distribution of average delays for the Noon scenario with lane disruptions at (a) 10:30 AM , and (b) 12:45 PM.}
    \label{fig:histograms}
\end{figure}

For the Noon scenario, deep RL still reduced delays, but the improvement is not as significant as that of the AM scenario (Fig. \ref{fig:AM_average}b). 
This is understandable for the Noon scenario as improvement is not as visible during off-peak periods. 
For the PM scenario, however, a significant improvement is seen only during the peak period (4:00 -- 7:00\,PM). 
There is a noticeable spike in delay at around 10:30\,AM and 6:30\,PM for the Noon and PM scenarios respectively.
Looking at the distribution of average delay values for the delay spike at 10:30\,AM, we see that the policy results in a wide spread of delay values (Fig. \ref{fig:histograms}a), which is not the case for the 6:30\,PM peak.
These peaks appear to be caused by the particular vehicle routes that are negatively impacted by the offset settings at those particular times, but we do note that even the baseline has this issue (11:45\,AM Fig. \ref{fig:AM_average}b).
The presence of these peaks does not appear to be caused by changes in control plans, or changes in offsets set by the RL policy.

\begin{table}[]
\caption{Average delays for different interval settings}
\centering
\begin{tabular}{ccccc}
& Interval & Baseline & Deep RL & \% Difference \\
\toprule
\multirow{ 5}{*}{AM} & 5-min    & 53.17  & 56.93 & -7.06 \\
& 10-min   & 53.13  & 55.87 & -5.15 \\
& 15-min   & 53.46  & 46.40 & 13.21 \\
& 30-min   & 53.17  & 54.03 & -1.60 \\
& 45-min   & 53.90  & 54.95  & -1.94\\
\midrule

\multirow{ 5}{*}{Noon}& 5-min    & 43.19  & 44.18 & -2.30 \\
& 10-min   & 43.16  & 44.01 & -1.97 \\
& 15-min   & 43.37  & 42.32 & 2.42  \\
& 30-min   & 43.23  & 43.90 & -1.55 \\
& 45-min   & 43.21  & 43.57 & -0.82 \\
\midrule

\multirow{ 5}{*}{PM} &5-min    & 45.18  & 47.34 & -4.79 \\
&10-min   & 45.16  & 46.57 & -3.13 \\
&15-min   & 45.89  & 43.04 & 6.20  \\
&30-min   & 45.19  & 46.88 & -3.74 \\
&45-min   & 45.60  & 46.55 & -2.09
\end{tabular}
\label{tab:intervals}
\end{table}

\begin{figure}[tb]
    \centering
    \includegraphics[width=3.3in]{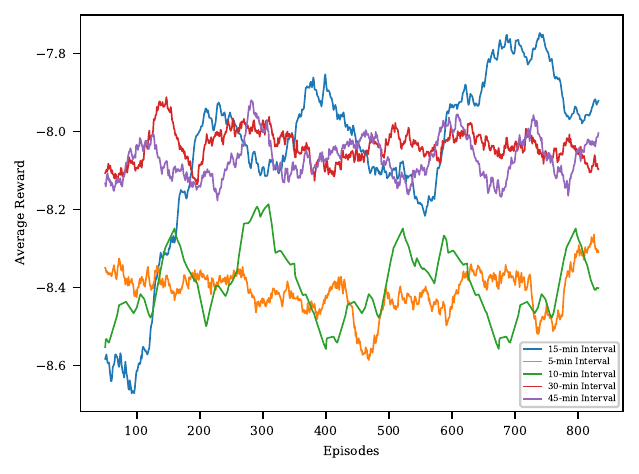}
    \caption{Average rewards (50 point moving averages) for different intervals. Highest rewards were obtained by the 15-minute interval, while the shorter intervals (5 and 10 minutes) reached much lower rewards than the longer intervals (30 and 45 minutes).}
    \label{fig:reward_intervals}
\end{figure}
While the results above were done for 15-minute intervals, we also explored other interval durations.
We trained separate policies using 5, 10, 30, and 45 minute intervals for 800 episodes (Fig. \ref{fig:reward_intervals}).
Sensor measurements were aggregated over these new intervals to compare the performance of the deep RL against the baselines (see Table \ref{tab:intervals})
For these intervals, the rewards attained using deep RL have much lower values for the shorter intervals (5 and 10 minutes) than that of the longer intervals (30 and 45 minutes).
Our model performs best for the 15-minute interval, while other interval settings fail to reduce the delays compared to our baseline.
Averaging over shorter intervals can result in more varied state observations, which delays learning.
On the other hand, averaging over longer intervals results in more stable detector measurements, but the agent fails to respond to changes in demand within the interval.
Other works change offsets at various intervals.
Adaptive algorithms adjust offsets periodically anywhere from 5-15 minutes \cite{gettman2007data}, as often as every cycle ($t_c=10\,\mathrm{sec}$) \cite{abbas2001real}, or after multiple cycles ($t_c=75\,\mathrm{sec}$, 10 cycles) \cite{zhang2015implementation}.
While we did not test for instabilities that may arise from changing offsets this frequently, our results for 15-minute intervals over multiple realizations suggest that the method is stable.

To test the agent's robustness, we evaluated the network performance of the learned policy under two exogenous uncertainties~\cite{rodrigues2019towards}.
First, we induced a surge in the demand by increasing the number of car trips by 50\% at off-peak times (specifically between 12 and 1\,PM).
Second, we induced disruptions by generating random traffic incidents (like a stalled car) that blocked a single lane on the first incoming Eastbound (Huntington and Santa Clara) and Westbound (Huntington and Gateway Drive) links entering the coordinated intersections over the period of 12\,NN to 1\,PM.
We introduced both of these exogenous uncertainties through the underlying Aimsun model.
Timing plans implemented in the field served as the performance baseline for the 70 realizations of the model for the Noon period. 

\begin{figure}[tb]
    \centering
    \includegraphics[width=3.2in]{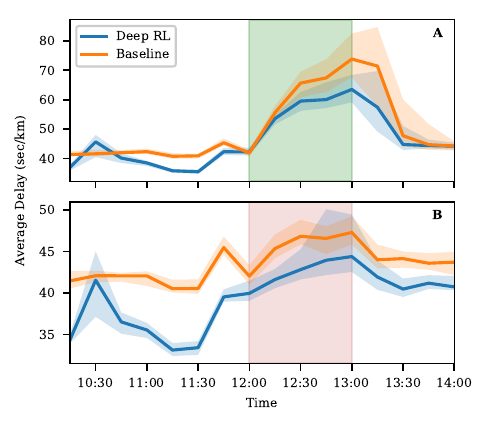}
    \caption{Average delay times for two different incidents (10\,AM -- 2\,PM): (a) Demand surge, and (b) Lane disruptions .
    Deep RL reduces the average delay times for both incidents compared to the baseline. }
    \label{fig:delay_time_interv}
\end{figure}

Figure \ref{fig:delay_time_interv} shows the average delay under increased demand (Green portion) and induced disruptions (Red portion). 
When the network is subjected to exogenous uncertainties, we observed reduced average delay times for the deep RL agent as compared to the baseline.
While the deep RL agent was not explicitly trained on traffic scenarios with disruptions, it surprisingly adapts its policy to the low flow, high occupancy states of the disrupted links.
However, there are certain instances where the average delays of RL policy have wide spreads which make them less effective than the baseline.
When exposed to lane disruptions, the RL agent's performance at 12:30\,PM can have delays that result in worse performance than the baseline, although the distribution still leans mostly to lower average delay values (Fig. \ref{fig:histograms}b).
Although we saw an overall improvement, the wider spread of delay for the deep RL results for lane disruptions shows that the baseline still remains more stable and predictable than deep RL.
However, in the case of traffic scenarios with demand surges, the RL policy is able to adapt and limit the increase in delays compared to the baseline.
While the baseline is not a dynamic algorithm (and hence not designed to adapt to changes in traffic conditions), this comparison provides a reference to the adaptability of the RL policy.

We also tested the deep RL agent against demand surges and lane disruptions for the peak times (AM and PM).
When subjected to demand surges during peak traffic periods, we did find similar and consistent results within the PM peak.
However, we found no significant difference between the average delays of deep RL and the baseline in the AM peak.
During the AM peak period, the traffic network is already heavily congested, and it would be difficult for any algorithm to handle an additional demand of 50\% more vehicles.
On the other hand, the agent can handle lane disruptions during peak hours, similar to the Noon scenario.
Traffic incidents pose a different challenge because the cause of congestion is localized to the affected links.
Here, offsets can still be adjusted to avoid excessive congestion in the affected links while maintaining the performance of the rest of the network.
These responses to exogenous uncertainties demonstrate our agent's robustness and flexibility.
We showed the agent generally adapts to extreme scenarios like demand surge and traffic incidents, despite not encountering such events during training.

\section{Conclusion}
\label{conclusion}
In this article, we proposed a framework that uses deep RL with Proximal Policy Optimization for traffic control coordination through dynamic adjustment of phase offsets. 
We utilized arterial traffic models adopted from the Connected Corridors program to train on a more realistic and non-synthetic arterial traffic environment. 

Our RL framework avoids the \emph{curse of dimensionality} through sensor-based state and reward representations and actions limited to only modify individual phase offsets, leaving the phase timings and order unmodified. 

We have proposed a framework that, to the best of our knowledge, has not been used before in other related deep RL for traffic control literature;
the phase splits remain model-based while the coordination is model free. 
Moreover, while most related works use value-based RL algorithms like DQN, our approach uses a policy-based algorithm, Proximal Policy Optimization (PPO). The result suggests that our deep RL approach beats the baseline in delay reduction even under exogenous uncertainties in the network. 

We have demonstrated that our deep RL agent design is viable for this coordination problem.
Simulation results are currently limited to one test site but can extend to other corridors provided a well-calibrated model is available.
In the future, we are interested in conducting an in-depth analysis of the model's sensitivity to different detector layouts and different levels of sensor degradation. We are also interested in integrating data from connected vehicles to improve the model's performance. In addition, we would like to extend our model to multi-agent RL to optimize the coordination in large-scale networks. In this case, a large-scale network can be subdivided into smaller intersection groups, each coordinated by its own RL agent.

Finally, we argue that while these deep RL agents have been proven to work in simulation, the importance of human intervention and engineering judgment must still govern over model-free approaches, especially in the context of traffic control where generated policies must be interpretable, regulatable, and reliable. This is the reason why we sought to find the balance between model-based and model-free traffic control.
While this is still premature for actual application, it is best to view this as a planning tool supplementary to decision-making processes.

\section*{Acknowledgments}
Keith Anshilo Diaz, Damian Dailisan, Carissa Santos, May Lim, Francis Aldrine Uy, and Alexandre Bayen would like to acknowledge the support  provided by the  Commission on Higher Education: Philippine--California Advanced Research Institutes [IIID-2016-006]. We also thank Victor Chan and members of the Mobile Sensing Lab for fruitful discussions.

\ifCLASSOPTIONcaptionsoff
 \newpage
\fi

\bibliographystyle{IEEEtran}
\bibliography{IEEEabrv,references}

\begin{thebibliography}{10}
\providecommand{\url}[1]{#1}
\csname url@samestyle\endcsname
\providecommand{\newblock}{\relax}
\providecommand{\bibinfo}[2]{#2}
\providecommand{\BIBentrySTDinterwordspacing}{\spaceskip=0pt\relax}
\providecommand{\BIBentryALTinterwordstretchfactor}{4}
\providecommand{\BIBentryALTinterwordspacing}{\spaceskip=\fontdimen2\font plus
\BIBentryALTinterwordstretchfactor\fontdimen3\font minus
  \fontdimen4\font\relax}
\providecommand{\BIBforeignlanguage}[2]{{%
\expandafter\ifx\csname l@#1\endcsname\relax
\typeout{** WARNING: IEEEtran.bst: No hyphenation pattern has been}%
\typeout{** loaded for the language `#1'. Using the pattern for}%
\typeout{** the default language instead.}%
\else
\language=\csname l@#1\endcsname
\fi
#2}}
\providecommand{\BIBdecl}{\relax}
\BIBdecl

\bibitem{cookson2017inrix}
G.~Cookson and B.~Pishue, ``Inrix global traffic scorecard--appendices,''
  \emph{INRIX research}, 2017.

\bibitem{ezell2010explaining}
S.~Ezell, ``Explaining international it application leadership: Intelligent
  transportation systems,'' 2010.

\bibitem{little1966synchronization}
J.~D. Little, ``The synchronization of traffic signals by mixed-integer linear
  programming,'' \emph{Operations Research}, vol.~14, no.~4, pp. 568--594,
  1966.

\bibitem{gartner1991multi}
N.~H. Gartner, S.~F. Assman, F.~Lasaga, and D.~L. Hou, ``A multi-band approach
  to arterial traffic signal optimization,'' \emph{Transportation Research Part
  B: Methodological}, vol.~25, no.~1, pp. 55--74, 1991.

\bibitem{robertson1969transyt}
D.~I. Robertson, ``Transyt: a traffic network study tool,'' 1969.

\bibitem{wang2018review}
Y.~Wang, X.~Yang, H.~Liang, and Y.~Liu, ``A review of the self-adaptive traffic
  signal control system based on future traffic environment,'' \emph{Journal of
  Advanced Transportation}, vol. 2018, 2018.

\bibitem{roess2004traffic}
R.~P. Roess, E.~S. Prassas, and W.~R. McShane, \emph{Traffic
  engineering}.\hskip 1em plus 0.5em minus 0.4em\relax Pearson/Prentice Hall,
  2004.

\bibitem{zou2004timing}
Z.-Y. Zou, S.-K. Chen, J.-y. GUO, L.-q. BAI, and C.-f. CHANG, ``Timing
  optimization and simulation on signalized intersection by synchro [j],''
  \emph{Journal of Northern Jiaotong University}, vol.~6, 2004.

\bibitem{gettman2007data}
D.~Gettman, S.~G. Shelby, L.~Head, D.~M. Bullock, and N.~Soyke, ``Data-driven
  algorithms for real-time adaptive tuning of offsets in coordinated traffic
  signal systems,'' \emph{Transportation Research Record}, vol. 2035, no.~1,
  pp. 1--9, 2007.

\bibitem{rodrigues2019towards}
F.~Rodrigues and C.~L. Azevedo, ``Towards robust deep reinforcement learning
  for traffic signal control: Demand surges, incidents and sensor failures,''
  in \emph{2019 IEEE Intelligent Transportation Systems Conference
  (ITSC)}.\hskip 1em plus 0.5em minus 0.4em\relax IEEE, 2019, pp. 3559--3566.

\bibitem{mnih2015human}
V.~Mnih, K.~Kavukcuoglu, D.~Silver, A.~A. Rusu, J.~Veness, M.~G. Bellemare,
  A.~Graves, M.~Riedmiller, A.~K. Fidjeland, G.~Ostrovski \emph{et~al.},
  ``Human-level control through deep reinforcement learning,'' \emph{Nature},
  vol. 518, no. 7540, pp. 529--533, 2015.

\bibitem{silver2017mastering}
D.~Silver, J.~Schrittwieser, K.~Simonyan, I.~Antonoglou, A.~Huang, A.~Guez,
  T.~Hubert, L.~Baker, M.~Lai, A.~Bolton \emph{et~al.}, ``Mastering the game of
  go without human knowledge,'' \emph{Nature}, vol. 550, no. 7676, pp.
  354--359, 2017.

\bibitem{Wei2018intellilight}
H.~Wei, H.~Yao, G.~Zheng, and Z.~Li, ``{IntelliLight: A reinforcement learning
  approach for intelligent traffic light control},'' \emph{Proc. ACM SIGKDD
  Int. Conf. Knowl. Discov. Data Min.}, pp. 2496--2505, 2018.

\bibitem{Wei2019presslight}
H.~Wei, C.~Chen, G.~Zheng, K.~Wu, V.~Gayah, K.~Xu, and Z.~Li, ``{Presslight:
  Learning Max pressure control to coordinate traffic signals in arterial
  network},'' pp. 1290--1298, 2019.

\bibitem{Wei2019colight}
\BIBentryALTinterwordspacing
H.~Wei, N.~Xu, H.~Zhang, G.~Zheng, X.~Zang, C.~Chen, W.~Zhang, Y.~Zhu, K.~Xu,
  and Z.~Li, ``{Colight: Learning network-level cooperation for traffic signal
  control},'' \emph{Int. Conf. Inf. Knowl. Manag. Proc.}, pp. 1913--1922, 2019.
  [Online]. Available: \url{http://arxiv.org/abs/1905.05717}
\BIBentrySTDinterwordspacing

\bibitem{belletti2017expert}
F.~Belletti, D.~Haziza, G.~Gomes, and A.~M. Bayen, ``Expert level control of
  ramp metering based on multi-task deep reinforcement learning,'' \emph{IEEE
  Transactions on Intelligent Transportation Systems}, vol.~19, no.~4, pp.
  1198--1207, 2017.

\bibitem{dion2015connected}
F.~Dion, ``Connected corridors: I-210 pilot integrated corridor management
  system concept of operations,'' \emph{California PATH, Berkeley, CA}, 2015.

\bibitem{gan2017estimation}
Q.~Gan, G.~Gomes, and A.~Bayen, ``Estimation of performance metrics at
  signalized intersections using loop detector data and probe travel times,''
  \emph{IEEE Transactions on Intelligent Transportation Systems}, vol.~18,
  no.~11, pp. 2939--2949, 2017.

\bibitem{gan2019arterial}
Q.~Gan and A.~Skabardonis, \emph{Arterial Traffic Estimation Using Field
  Detector and Signal Phasing Data}.\hskip 1em plus 0.5em minus 0.4em\relax UC
  Office of the President: University of California Institute of Transportation
  Studies, 2019.

\bibitem{bayen2011mobile}
A.~M. Bayen, A.~D. Patire, and J.~Butler, \emph{Mobile Millennium final
  report}.\hskip 1em plus 0.5em minus 0.4em\relax California Center for
  Innovative Transportation, Institute of Transportation Studies, University of
  California, Berkeley, 2011.

\bibitem{webster1958traffic}
F.~V. Webster, ``Traffic signal settings,'' Tech. Rep., 1958.

\bibitem{dunne1964algorithm}
M.~C. Dunne and R.~B. Potts, ``Algorithm for traffic control,''
  \emph{Operations Research}, vol.~12, no.~6, pp. 870--881, 1964.

\bibitem{wei2019survey}
H.~Wei, G.~Zheng, V.~Gayah, and Z.~Li, ``A survey on traffic signal control
  methods,'' \emph{arXiv preprint arXiv:1904.08117}, 2019.

\bibitem{park2010quantifying}
B.~B. Park and Y.~Chen, ``Quantifying the benefits of coordinated actuated
  traffic signal systems: A case study,'' Tech. Rep., 2010.

\bibitem{coogan2017offset}
S.~Coogan, E.~Kim, G.~Gomes, M.~Arcak, and P.~Varaiya, ``Offset optimization in
  signalized traffic networks via semidefinite relaxation,''
  \emph{Transportation Research Part B: Methodological}, vol. 100, pp. 82--92,
  2017.

\bibitem{yin2007offline}
Y.~Yin, M.~Li, and A.~Skabardonis, ``Offline offset refiner for coordinated
  actuated signal control systems,'' \emph{Journal of transportation
  engineering}, vol. 133, no.~7, pp. 423--432, 2007.

\bibitem{abbas2001real}
M.~Abbas, D.~Bullock, and L.~Head, ``Real-time offset transitioning algorithm
  for coordinating traffic signals,'' \emph{Transportation Research Record},
  vol. 1748, no.~1, pp. 26--39, 2001.

\bibitem{wu2017flow}
C.~Wu, A.~Kreidieh, K.~Parvate, E.~Vinitsky, and A.~M. Bayen, ``Flow:
  Architecture and benchmarking for reinforcement learning in traffic
  control,'' \emph{arXiv preprint arXiv:1710.05465}, 2017.

\bibitem{vinitsky2018benchmarks}
E.~Vinitsky, A.~Kreidieh, L.~Le~Flem, N.~Kheterpal, K.~Jang, C.~Wu, F.~Wu,
  R.~Liaw, E.~Liang, and A.~M. Bayen, ``Benchmarks for reinforcement learning
  in mixed-autonomy traffic,'' in \emph{Conference on Robot Learning}, 2018,
  pp. 399--409.

\bibitem{van2016coordinated}
E.~Van~der Pol and F.~A. Oliehoek, ``Coordinated deep reinforcement learners
  for traffic light control,'' \emph{Proceedings of Learning, Inference and
  Control of Multi-Agent Systems (at NIPS 2016)}, 2016.

\bibitem{schultz2018deep}
L.~Schultz and V.~Sokolov, ``Deep reinforcement learning for dynamic urban
  transportation problems,'' \emph{arXiv preprint arXiv:1806.05310}, 2018.

\bibitem{Casas2010}
\BIBentryALTinterwordspacing
J.~Casas, J.~L. Ferrer, D.~Garcia, J.~Perarnau, and A.~Torday, \emph{{Traffic
  Simulation with Aimsun}}.\hskip 1em plus 0.5em minus 0.4em\relax New York,
  NY: Springer New York, 2010, pp. 173--232. [Online]. Available:
  \url{https://doi.org/10.1007/978-1-4419-6142-6_5}
\BIBentrySTDinterwordspacing

\bibitem{SUMO2018}
P.~A. Lopez, M.~Behrisch, L.~Bieker-Walz, J.~Erdmann, Y.-P.
  Fl{\"{o}}tter{\"{o}}d, R.~Hilbrich, L.~L{\"{u}}cken, J.~Rummel, P.~Wagner,
  and E.~Wie{\ss}ner, ``{Microscopic Traffic Simulation using SUMO},'' in
  \emph{IEEE Intell. Transp. Syst. Conf.}\hskip 1em plus 0.5em minus
  0.4em\relax IEEE, 2018.

\bibitem{el2014design}
S.~El-Tantawy, B.~Abdulhai, and H.~Abdelgawad, ``Design of reinforcement
  learning parameters for seamless application of adaptive traffic signal
  control,'' \emph{Journal of Intelligent Transportation Systems}, vol.~18,
  no.~3, pp. 227--245, 2014.

\bibitem{ge2019cooperative}
H.~Ge, Y.~Song, C.~Wu, J.~Ren, and G.~Tan, ``Cooperative deep q-learning with
  q-value transfer for multi-intersection signal control,'' \emph{IEEE Access},
  vol.~7, pp. 40\,797--40\,809, 2019.

\bibitem{chen2019adaptive}
P.~Chen, Z.~Zhu, and G.~Lu, ``An adaptive control method for arterial signal
  coordination based on deep reinforcement learning,'' in \emph{2019 IEEE
  Intelligent Transportation Systems Conference (ITSC)}.\hskip 1em plus 0.5em
  minus 0.4em\relax IEEE, 2019, pp. 3553--3558.

\bibitem{wiering2000multi}
M.~Wiering, ``Multi-agent reinforcement learning for traffic light control,''
  in \emph{Machine Learning: Proceedings of the Seventeenth International
  Conference (ICML'2000)}, 2000, pp. 1151--1158.

\bibitem{mikami1994genetic}
S.~Mikami and Y.~Kakazu, ``Genetic reinforcement learning for cooperative
  traffic signal control,'' in \emph{Proceedings of the First IEEE Conference
  on Evolutionary Computation. IEEE World Congress on Computational
  Intelligence}.\hskip 1em plus 0.5em minus 0.4em\relax IEEE, 1994, pp.
  223--228.

\bibitem{bucsoniu2010multi}
L.~Bu{\c{s}}oniu, R.~Babu{\v{s}}ka, and B.~De~Schutter, ``Multi-agent
  reinforcement learning: An overview,'' in \emph{Innovations in multi-agent
  systems and applications-1}.\hskip 1em plus 0.5em minus 0.4em\relax Springer,
  2010, pp. 183--221.

\bibitem{steingrover2005reinforcement}
M.~Steingrover, R.~Schouten, S.~Peelen, E.~Nijhuis, B.~Bakker \emph{et~al.},
  ``Reinforcement learning of traffic light controllers adapting to traffic
  congestion.'' in \emph{BNAIC}.\hskip 1em plus 0.5em minus 0.4em\relax
  Citeseer, 2005, pp. 216--223.

\bibitem{el2013multiagent}
S.~El-Tantawy, B.~Abdulhai, and H.~Abdelgawad, ``{Multiagent reinforcement
  learning for integrated network of adaptive traffic signal controllers
  (MARLIN-ATSC): methodology and large-scale application on downtown
  Toronto},'' \emph{IEEE Transactions on Intelligent Transportation Systems},
  vol.~14, no.~3, pp. 1140--1150, 2013.

\bibitem{houli2010multiobjective}
D.~Houli, L.~Zhiheng, and Z.~Yi, ``Multiobjective reinforcement learning for
  traffic signal control using vehicular ad hoc network,'' \emph{EURASIP
  journal on advances in signal processing}, vol. 2010, no.~1, p. 724035, 2010.

\bibitem{dulac2015deep}
G.~Dulac-Arnold, R.~Evans, H.~van Hasselt, P.~Sunehag, T.~Lillicrap, J.~Hunt,
  T.~Mann, T.~Weber, T.~Degris, and B.~Coppin, ``Deep reinforcement learning in
  large discrete action spaces,'' \emph{arXiv preprint arXiv:1512.07679}, 2015.

\bibitem{kuyer2008multiagent}
L.~Kuyer, S.~Whiteson, B.~Bakker, and N.~Vlassis, ``Multiagent reinforcement
  learning for urban traffic control using coordination graphs,'' in
  \emph{Joint European Conference on Machine Learning and Knowledge Discovery
  in Databases}.\hskip 1em plus 0.5em minus 0.4em\relax Springer, 2008, pp.
  656--671.

\bibitem{sutton1998introduction}
R.~S. Sutton, A.~G. Barto \emph{et~al.}, \emph{Introduction to reinforcement
  learning}.\hskip 1em plus 0.5em minus 0.4em\relax MIT press Cambridge, 1998,
  vol. 135.

\bibitem{bellman1957markovian}
R.~Bellman, ``A markovian decision process,'' \emph{Journal of mathematics and
  mechanics}, pp. 679--684, 1957.

\bibitem{schulman2017proximal}
J.~Schulman, F.~Wolski, P.~Dhariwal, A.~Radford, and O.~Klimov, ``Proximal
  policy optimization algorithms,'' 2017.

\bibitem{Zheng2019}
G.~Zheng, X.~Zang, N.~Xu, H.~Wei, Z.~Yu, V.~Gayah, K.~Xu, and Z.~Li,
  ``{Diagnosing Reinforcement Learning for Traffic Signal Control},'' 2019.

\bibitem{haydari2020deep}
A.~{Haydari} and Y.~{Yilmaz}, ``Deep reinforcement learning for intelligent
  transportation systems: A survey,'' \emph{IEEE Transactions on Intelligent
  Transportation Systems}, pp. 1--22, 2020.

\bibitem{Schulman}
J.~Schulman, P.~Moritz, S.~Levine, M.~Jordan, and P.~Abbeel, ``High-dimensional
  continuous control using generalized advantage estimation,'' \emph{arXiv
  preprint arXiv:1506.02438}, 2015.

\bibitem{zhang2015implementation}
L.~Zhang, Z.~Huang, Y.~Wen, A.~Hawkins, W.~C. Fulcher, S.~R. Henke, A.~Roberts,
  and Q.~Zhang, ``Implementation of real-time offset-tuning algorithm for
  integrated corridor management,'' \emph{Transportation Research Record}, vol.
  2488, no.~1, pp. 97--107, 2015.

\end{thebibliography}

\end{document}